\documentclass[twocolumn,amsmath,showkeys,amssymb,superscriptaddress,longbibliography,floatfix, pre]{revtex4-2}

\pdfoutput=1

\usepackage{bm}
\usepackage{yfonts}
\usepackage{graphicx,epsfig}
\usepackage{amsmath,amssymb,mathtools}
\usepackage[rgb]{xcolor}
\usepackage{hyperref}
\usepackage{braket}
\def\be{\begin{equation}}
\def\ee{\end{equation}}

\def\bc{\begin{center}}
\def\ec{\end{center}}
\def\bea{\begin{eqnarray}}
\def\eea{\end{eqnarray}}

\begin{document}

\title{The  quantum relative entropy of the Schwarzschild  black-hole and the area law}

\author{Ginestra Bianconi}
\email{ginestra.bianconi@gmail.com}
\affiliation{School of Mathematical Sciences, Queen Mary University of London, London, E1 4NS, United Kingdom}

\begin{abstract}
The area law obeyed by the thermodynamic entropy of black holes is one of the fundamental results relating gravity to statistical mechanics.
In this work we provide a derivation of the area law for the quantum relative entropy of the Schwarzschild black-hole for arbitrary Schwarzschild radius. The quantum relative entropy between the metric of the manifold and the  metric induced by the geometry and the matter field has been  proposed in G. Bianconi {\em Gravity from entropy}, Phys. Rev. D (2025) as the action for  entropic quantum gravity leading to modified Einstein equations. The quantum relative entropy generalizes Araki entropy and  treats the metrics between zero-forms, one-forms, and two-forms as quantum operators.
Although the Schwarzschild  metric is not an exact solution of the modified Einstein equations of the entropic quantum gravity,  it is  an approximate solution valid  in the low coupling, small curvature limit.
Here we show that the quantum relative entropy associated to the Schwarzschild  metric obeys the area law for large Schwarzschild radius. We provide a full statistical mechanics interpretation of the results.
\end{abstract}
\maketitle

 \section{Introduction}
The area law satisfied by the thermodynamic entropy of black holes is one of the cornerstones of quantum gravity~\cite{witten2024introduction,mukhanov2007introduction,wald1994quantum}. The discovery that  the entropy, notoriously an extensive quantity, can obey an area law came as a big surprise of the early findings of  Bekenstein~\cite{bekenstein1973black,bekenstein1974generalized} and Hawking \cite{hawking1975particle,gibbons1977action} and continues to stimulate theoretical physics explanations. Indeed, after the discovery of this law, the study of the entropy of black holes~\cite{wald1993black,calmet2021quantum} and the area law became a testbench for quantum gravity approaches leading to explanations making use of string theory \cite{strominger1996microscopic}, the holographic principle \cite{hooft1993dimensional,susskind1995world}, the AdS/CFT correspondence \cite{maldacena1999large}, and in particular the Ryu--Takayanagi formula \cite{ryu2006aspects} and loop quantum gravity approaches \cite{rovelli1996black}.
The area law is also considered a universal property of condensed matter systems \cite{amico2008entanglement,eisert2010colloquium} as it has an important interpretation in terms of the entanglement entropy. Interestingly, recent approaches have provided new insights into black-hole entropy by quantifying the entanglement entropy of scalar fields near the horizon of black-holes~\cite{Luongo1,Luongo2}.

In this work, we discuss the quantum relative entropy of the Schwarzschild black hole. The quantum relative entropy  is a fundamental information theory quantity \cite{vedral2002role} whose importance is central in quantum information and the theory of quantum operators~\cite{araki1975relative,araki1999mathematical,witten2018aps,ohya2004quantum}. Recently, in Ref.~\cite{bianconi2024gravity}, the quantum relative entropy has been proposed by G. Bianconi  as the fundamental information theory action for the entropic quantum gravity  approach.
The definition of the quantum relative entropy relies on the treatment of the spacetime metric and the metric induced by the geometry of spacetime and the matter fields as quantum operators.
Note that the idea that the considered manifold is described by two metrics is at the foundation of the bi-metric gravitation~\cite{rosen1973bi,hossenfelder2008bimetric} as well. However, the treatment of these two metrics as quantum operators and the use of  the quantum relative entropy between the two metrics as the action  for gravitation make the entropic quantum gravity approach significant distinct from the bi-metric approach.  In the entropic quantum gravity approach,  the action for gravity is the  quantum relative entropy between the metric of the considered manifold and the metric induced by the geometry and the matter field. A fundamental aspect of the entropic quantum gravity approach is that the two considered metrics are topological, i.e., they are the direct sum of metrics between zero-forms, one-forms, and two-forms. Thus, this aspect of the entropic quantum approach is in line with growing interest in area metrics in quantum gravity
\cite{schuller2006geometry,borissova2024area,kuipers2024quantum,dittrich2023spin}. 

The entropic quantum gravity  approach leads to modified Einstein equations, which reduce to the Einstein equations in the low-coupling, small-curvature limit. 
However, the action of the entropic quantum gravity is very different from the Einstein--Hilbert action. Among the important differences, we observe that, thanks to the inclusion of metrics between two-forms,  the  entropic quantum gravity action depends explicitly on the Riemann tensor; therefore, it is not vanishing for a Schwarzschild  black hole.

In this work, we perform a derivation of the area law for the quantum relative entropy associated to the Schwarzschild  black hole. It is to be noted that the Schwarzschild  metric is not an exact black-hole solution of the entropic quantum gravity approach; however, it is a solution in the low-coupling, small-curvature regime.
The area law is recovered exactly in this limit, i.e., when the Schwarzschild  radius is very large, although the multiplicative constants are different than the ones predicted for the thermodynamic entropy. Moreover,  for a small radius, deviations from the area law are observed.

Recently, we have entered  a phase of experimental tests of gravity combining results coming from different experimental sources. This includes, of course, the validation coming from gravitational wave experiments~\cite{berti2015testing,isi2021testing} and also includes validations of analogue gravity~\cite{barcelo2001analogue,weinfurtner2011measurement,gooding2020interferometric} and the exploration of gravity effects by the means of quantum information theory
\cite{bose2017spin,marletto2017gravitationally,marletto2024quantum,howl2021non}.
Thus, it is our hope that these results can contribute to providing experimental probes of the quantum gravity effects in nature.

\section{Boltzmann Legacy and the Quantum Relative Entropy for Gravity}
\begin{figure*}[!htb]
  \includegraphics[width=0.9\textwidth]{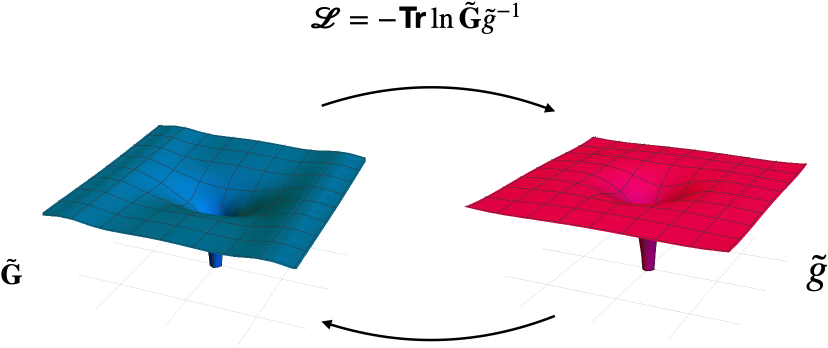}
  \caption{ {Diagrammatic} 
 sketch of the entropic quantum gravity approach. In this approach the action  is given by the  quantum relative entropy between the metric $\tilde{g}$ and the metric $\tilde{\bf G}$ induced by the matter fields and the geometry of the  manifold.}
  \label{fig_sketch}
 \end{figure*}

Our starting point is the celebrated expression for the entropy given by Boltzmann~\cite{huang1987introduction}, which provides a microscopic interpretation of the thermodynamics entropy\linebreak $S=S(E,V)$ for a system of given total energy $E$ and volume $V$, i.e.,
\bea
S(E,V)=k_B\ln W
\eea
Here, $k_B$ indicates the Boltzmann constant, while $W$ indicates the number of microstate configurations compatible with the considered macrostate configuration.
One classical result of this formula is that the entropy is extensive. This implies that for a system of locally interacting set of $N$ identical particles in thermal equilibrium, which can be considered as the sum of two systems at the same temperature (a system of $N_1$ and a system of $N_2$ particles), the total number of particles $N$, the total volume $V$,  and the total energy $E$ of the system can be written as
\bea
N=N_1+N_2,\nonumber \\
V=V_1+V_2,\nonumber \\
E=E_1+E_2,
\eea
where $V_i$ and $E_i$ for $i\in \{1,2\}$ are the volume and the energy of the two subsystems, respectively. In this scenario, we have that $W$ obeys
\bea
\ln W=\ln W_1+\ln W_2 +O(\ln N),
\eea
where $W_i$ is the number of microscopic configurations compatible with the subsystem $i$.
It follows that the entropy $S(E,V)$ is extensive, i.e.,
\bea
S(E,V)=S_1(E_1,V_1)+S_2(E_2,V_2)+O(\ln N).
\eea
Thus, considering a system as composed by $n$ macroscopic subsystems $i\in \{1,2,\ldots,n\}$, we~obtain
\bea
S(E,V)=k_B\sum_{i=1}^n\ln W_i.
\label{ext}
\eea
In gravity, the degrees of freedom are encoded in the spacetime fabric of a $d=4$ dimensional manifold $\mathcal{K}$  of Lorentzian signature $\{-1,1,1,1\}$, whose geometry is fully described by its metric $g_{\mu\nu}$.
In the entropic quantum gravity proposed in Ref. \cite{bianconi2024gravity}, the topological metric considered comprises the  metric among scalars, the metric among vectors, and the metric among bivectors defined in $\mathcal{K}$.
This is given by 
\bea
\tilde{g}&=&1 \oplus g_{\mu\nu} dx^{\nu}\otimes dx^{\nu} \\
&&\oplus {[g_{(2)}]}_{\mu\nu\rho\sigma}(dx^{\mu}\wedge dx^{\nu})\otimes(dx^{\rho}\wedge dx^{\sigma}).\nonumber
\eea
where 
\bea
{[g_{(2)}]}_{\mu\nu\rho\sigma}=\frac{1}{2}(g_{\mu\rho}g_{\nu\sigma}-g_{\mu\sigma}g_{\nu\rho}).
\label{g2}
\eea
Additionally the topological metric $\tilde{\bf G}$ induced by the geometry and the matter fields is also considered; this metric also comprises the direct sum between a metric among scalars $\tilde{G}_{(0)}$, a metric among vectors $\tilde{G}_{(0)}$, and a metric among bivectors $\tilde{G}_{(1)}$ and is  given by 
\bea
\tilde{\bf G}&=&\tilde{G}_{(0)}\oplus[\tilde{G}_{(1)}]_{\mu\nu}dx^{\mu}\otimes dx^{\nu}\nonumber \\
&&\oplus [\tilde{G}_{(2)}]_{\mu\nu\rho\sigma}(dx^{\mu}\wedge dx^{\nu})\otimes(dx^{\rho}\wedge dx^{\sigma}),
\eea
where at each point $p$ of the manifold $\mathcal{K}$, the matrices $\tilde{\bf G}_{(m)}$ with $m\in \{0,1,2\}$ are invertible.
The  dual metric is given by $\tilde{\bf G}^{\star}$
\bea
\tilde{\bf G}^{\star}&=&\tilde{G}_{(0)}\oplus[\tilde{G}_{(1)}]^{\mu\nu}dx_{\mu}\otimes dx_{\nu}\nonumber\\
&&\oplus [\tilde{G}_{(2)}]^{\mu\nu\rho\sigma}(dx_{\mu}\wedge dx_{\nu})\otimes(dx_{\rho}\wedge dx_{\sigma}).
\eea
The entropic quantum gravity approach proposed in Ref. \cite{bianconi2024gravity} considers the following entropic action for modified gravity given by the quantum relative entropy between $\tilde{\bf G}$ and $g$ (see Figure $\ref{fig_sketch}$ for a diagrammatic description), i.e.,
\bea
\mathcal{S}=\frac{1}{\ell_P^d}\int \sqrt{|-{g}|}\mathcal{L} d{\bf r},
\label{S}\eea
where $\ell_P=(\hbar G/c^3)^{1/2}$ is the Planck length, and the Lagrangian is given by
\bea
\mathcal{L}&:=&-{\mbox{Tr}}\ln \tilde{\bf G}{\tilde{ g}^{-1}}\nonumber \\
&:=&-\ln \tilde{G}_{(0)}-{\mbox{Tr}}\ln \tilde{\bf G}_{(1)}{{ g}^{-1}}-{\mbox{Tr}}\ln {\tilde{\bf G}_{(2)}}{g}_{(2)}^{-1},
\label{L0}
\eea
where here and in the following by $\mbox{Tr}$ we always imply the flattened trace defined in Ref. \cite{bianconi2024gravity} and indicated there as $\mbox{Tr}_F$.
Here, we assume that $\tilde{\bf G}\tilde{g}^{-1}$ is positively defined, i.e.,  $\tilde{G}_{(0)}>0$ and $\tilde{\bf G}_{(1)}g^{-1}$, as well as $\tilde{\bf G}_{(2)}g_{(2)}^{-1}$, are positively defined at each point $p$ of the manifold $\mathcal{K}$.
Note that this entropic action is expressed in terms of the square root of the modular operator $\boldmath{\Delta}_{\tilde{\bf G},\tilde{g}}
^{1/2}$ given by
\bea
\tilde{\bf G}\tilde{g}^{-1}=\boldmath{\Delta}_{\tilde{\bf G},\tilde{g}}
^{1/2}=\sqrt{\tilde{\bf G}\tilde{\bf G}^{\star}},\eea
thus generalizing the definition of the Araki entropy \cite{araki1975relative} between quantum operators to the considered topological metrics $\tilde{g}$ and $\tilde{\bf G}$ (see Ref. \cite{bianconi2024gravity} for a more detailed discussion).

We observe that the action for the entropic quantum gravity approach also admits an information theory interpretation akin to the Boltzmann entropy.
In fact, we have that the Lagrangian $\mathcal{L}$ can be written as 
\bea
\mathcal{L}=-\mbox{Tr}\ln \tilde{\bf G}\tilde{g}^{-1}=\ln W({\bf r}),
\eea 
where $W({\bf r})$ ``counts''  the degrees of freedom of the geometry, albeit it is in general a real rather than an integer number. In particular we have 
\bea
W({\bf r})=\tilde{G}_{(0)}^{-1}\det(\tilde{{\bf G}}_{(1)}^{-1}g)\det(\tilde{{\bf G}}_{(2)}^{-1}g_{(2)}).
\eea
Consequently the quantum relative entropy $\mathcal{S}$ can be written in a way reminiscent of Equation (\ref{ext})  as
\bea
\mathcal{S}=\frac{1}{\ell_P^4}\int \sqrt{-|g|} \ln W({\bf r}) d{\bf r}.
\eea
Thus, the quantum relative entropy counts the number of degrees of freedom of the metric and is  associated with the volume over which the integral is performed.

\section{Modified Einstein Equations in Vacuum}
The entropic quantum gravity approach leads to modified Einstein equations, which reduce to the Einstein equation in a regime of low coupling (small curvature and low energies).
Here, we are interested in discussing the  corresponding modified Einstein equations in vacuum and showing that the Schwarzschild  solutions are approximate solutions of these modified Einstein equations in the low-coupling regime.
In vacuum,  adopting the units $\hbar=c=1$, the expression of the metric induced by the geometry is assumed (see Ref. \cite{bianconi2024gravity}) to be given by 
\bea
\tilde{\bf G}=\tilde{g}-\beta G\tilde{\bm{\mathcal{R}}},
\label{Gv}
\eea
where $G$ is the gravitational constant, $\beta$ is a adimensional constant and $\tilde{\bm{\mathcal{R}}}$ is given by the topological curvature, comprising the Ricci scalar $R$, the Ricci tensor $R_{\mu\nu}$, and the Riemann tensor $R_{\mu\nu\rho\sigma}$, i.e.,
\bea
\tilde{\bm{\mathcal{R}}}&=&R\oplus \Big(R_{\mu\nu} dx^{\mu}\otimes dx^{\nu}\Big)\nonumber \\&&\oplus R_{\mu\nu\rho\sigma}(dx^{\mu}\wedge dx^{\nu})\otimes(dx^{\rho}\wedge dx^{\sigma}).
\eea
Leaving the discussion of the derivation of the modified Einstein equations derived from the entropic quantum gravity action $\mathcal{S}$ to Ref. \cite{bianconi2024gravity}, here, we summarize their structure.
The modified Einstein equations of entropic quantum gravity involve two sets of equations: the equations for the metric $\tilde{g}$ and the equations for the auxiliary G-fields $\tilde{\bm{\mathcal{G}}}$ (a form of auxiliary metric as well). 

 The equations for the G-fields $\tilde{\bm{\mathcal{G}}}$ are given by 
\bea
\tilde{\bm{\mathcal{G}}}^{-1}=\tilde{\bf I}-\beta G\tilde{\bm{\mathcal{R}}}\tilde{g}^{-1},
\label{Gm1b}
\eea
where $\tilde{\bm{\mathcal{G}}}^{-1}$ is the topological metric comprising a metric among scalars, one among vectors, and one among bivectors, each equal to the inverse of the corresponding metrics forming the topological metric $\tilde{\bm{\mathcal{G}}}$, and  $\tilde{\bf I}$ is the topological identity metric.

The modified Einstein equations for the metric $\tilde{g}$ are given by 
\bea
{R}^{\mathcal{G}}_{(\mu\nu)}-\frac{1}{2}{g}_{\mu\nu}\Big(\mathcal{R}_{\mathcal{G}}-2\Lambda_{\mathcal{G}}\Big)+{{\mathcal{D}}}_{(\mu\nu)}=0,
\label{modEin}
\eea
where 
\bea
\mathcal{R}_{\mathcal{G}}&=&\mbox{Tr}_{F}\tilde{g}_{\mathcal{G}}^{-1}\tilde{\bm{\mathcal{R}}},\nonumber \\
\Lambda_{\mathcal{G}}&=&\frac{1}{2\beta G}\mbox{Tr}_{F}\Big(\tilde{\bm{\mathcal{G}}}-\tilde{\bf I}-\ln \tilde{\bm{\mathcal{G}}}\Big),
\label{dressed_eq}
\eea
with $\tilde{g}_{\mathcal{G}}$ indicating a ``dressed metric'' given by
 \bea
 \tilde{g}_{\mathcal{G}}=\tilde{\bm{\mathcal{G}}}^{-1}g.
\label{gG} 
 \eea
Note that in Equation (\ref{modEin}),  $(\mu\nu)$ indicates the symmetrization of the indices, ${R}^{\mathcal{G}}_{\mu\nu}$ are the elements or the {\em dressed Ricci tensor} given by
\bea
{R}^{\mathcal{G}}_{\mu\nu}&=&{\mathcal{G}_{(0)}}R_{\mu\nu}+{[{\mathcal{G}_{(1)}}]}_{\mu}^{\ \rho}R_{\rho\nu}-{[\mathcal{G}_{(2)}]}_{\rho_1\rho_2\mu\eta}R_{\nu}^{\ \eta\rho_1\rho_2}\nonumber \\
&&+2{[{\mathcal{G}_{(2)}}]}_{\mu}^{\ \eta\rho_1\rho_2}R_{\rho_1\rho_2\nu\eta},
\eea
while ${{\mathcal{D}}}_{\mu\nu}$ are the elements depending on second derivatives of the G-field $\tilde{\bm{\mathcal{G}}}$ given by 
\bea
{{\mathcal{D}}}_{\mu\nu}&=&(\nabla^{\rho}\nabla_{\rho}g_{\mu\nu}-\nabla_{\mu}\nabla_{\nu}){\mathcal{G}_{(0)}}-\nabla^{\rho}\nabla_{\nu}{[\mathcal{G}_{(1)}]}_{(\rho\mu)}
\nonumber \\
&&+\frac{1}{2}
\nabla^{\rho}\nabla_{\rho}{[\mathcal{G}_{(1)}]}_{\mu\nu}+\frac{1}{2}\nabla^{\rho}\nabla^{\eta}{[\mathcal{G}_{(1)}]}_{\rho\eta}g_{\mu\nu}\nonumber \\
&&+\nabla^{\eta}\nabla^{\rho}{[\mathcal{G}_{(2)}]}_{\mu\rho\nu\eta}+\nabla^{\rho}\nabla^{\eta}{[\mathcal{G}_{(2)}]}_{\eta\mu\rho\nu}\nonumber \\
&&+\frac{1}{2}[\nabla^{\rho},\nabla^{\eta}]{[\mathcal{G}_{(2)}]}_{\rho\eta\mu\nu}.
\eea
These modified Einstein equations reduce to the Einstein equations in vacuum 
\bea
R_{\mu\nu}=0,
\eea
only if 
\bea
\tilde{\bm{\mathcal{G}}}^{-1}=\tilde{\bf I},
\eea 
i.e., only if 
\bea
\tilde{\bm{\mathcal{R}}}\tilde{g}^{-1}=0.\eea
However, the Einstein equations in vacuum remain a good approximation of the modified Einstein equations as long as 
\bea
\beta G\tilde{\bm{\mathcal{R}}}\tilde{g}^{-1}\ll
 \tilde{\bf I}.
\eea
Thus, the Schwarzschild  metric can be interpreted only as an  approximate solution of the modified Einstein equations of entropic quantum gravity in vacuum valid in the regime of small curvature. This implies that if the entropic quantum gravity approach captures the physics of gravitation, the physical black holes will  only be  described by the Schwarzschild metric in a linear approximation valid in the small-curvature regime.
  
Relevantly, however, we observe that  the quantum relative entropy $\mathcal{S}$  between $\tilde{\bf G}$  and $\tilde{g}$, given by Equation (\ref{Gv}) is defined for any metric, not only for the metric satisfying the mentioned equations for modified gravity. 
 Thus, in the next section, we address the challenge of evaluating  the quantum relative entropy $\mathcal{S}$ of the Schwarzschild metric.

\section{Quantum Relative Entropy of the Schwarzschild  Black Hole}
In this section, our goal is to provide the derivation of the quantum relative entropy of the Schwarzschild  black hole.
In particular, we show that the quantum relative entropy of the Schwarzschild  black hole follows an area law for large Schwarzschild  radii.
The starting point is the observation that the quantum relative entropy defining the entropic quantum gravity approach is not vanishing for a Schwarzschild  black hole as it depends explicitly on the Riemann tensor and not just exclusively on the Ricci scalar and the Ricci tensor.
This allows us to directly calculate the quantum relative entropy of the Schwarzschild black hole as a function of its Schwarzschild radius.

The Schwarzschild black hole defines the static and spherically symmetric metric 
\bea
\hspace{-5mm}ds^2=-\left(1-\frac{R_s}{r}\right)dt^2+{\left(1-\frac{R_s}{r}\right)}^{-1}dr^2+r^2d\Omega^2,
\label{SW}
\eea
where, in units $\hbar=c=1$, $R_s$ defines the Schwarzschild radius given by
\bea
R_s={2GM}, 
\eea
and where $d\Omega^2=d\theta^2+\sin^2\theta d\phi^2$. This  is the unique static and spherically symmetric metric  solution to the Einstein equations 
\bea
R_{\mu\nu}=0.
\eea
As discussed in the previous section, however, this is only the approximate solution to the entropic quantum gravity equations in vacuum, valid for small curvatures.

As mentioned before, the goal here is to calculate the quantum relative entropy $\mathcal{S}$ defined in Equations (\ref{S}) and (\ref{L0})  for the Schwarzschild metric defined in Equation (\ref{SW}).
To this end, we calculate explicitly the product between the metric induced by the geometry $\tilde{\bf G}$ in vacuum (Equation (\ref{Gv})) and the topological metric $\tilde{g}^{-1}$, i.e.,
\bea
\tilde{\bf G}\tilde{g}^{-1}=\tilde{\bf I}-\beta G\tilde{\bm{\mathcal{R}}}\tilde{g}^{-1}.\eea
By performing this straightforward calculation, we obtain
\bea
\hspace{-5mm}\tilde{\bf G}g^{-1}=1\oplus \delta_{\mu}^{\nu}dx^{\mu}\otimes dx_{\nu}+\Delta_{\mu\nu}^{\ \ \rho\sigma} dx^{\mu}\wedge dx^{\nu}\otimes dx_{\rho}\wedge dx_{\sigma},\nonumber
\eea
where
\bea
\Delta_{\mu\nu}^{\ \ \rho\sigma}=\frac{1}{2}(\delta_\mu^{\rho}\delta_{\nu}^{\sigma}-\delta_{\mu}^{\sigma}\delta_{\nu}^{\rho})-\beta G R_{\mu\nu}^{\ \ \rho\sigma}.
\eea
We then calculate the non-zero elements of the Riemann tensor $R_{\mu\nu}^{\ \ \rho\sigma}$ associated to the Schwarzschild metric to be 
\bea
R_{tr}^{\ \ tr}&=&R_{\theta\phi}^{\ \ \theta\phi}=\frac{R_s}{r^3},\nonumber \\
R_{t\theta}^{\ \ t\theta}&=&R_{t\phi}^{\ \ t\phi}=R_{r\theta}^{\ \ r\theta}=R_{r\phi}^{\ \ r\phi}=-\frac{R_s}{2r^3}.
\eea
Inserting these expressions in the Lagrangian $\mathcal{L}$ defined in Equation (\ref{L0}), we obtain
\bea
\hspace{-7mm}\mathcal{L}=-\mbox{Tr} \ln \Delta =-\ln \left[{\left(1-2\beta\frac{GR_s}{r^3}\right)}^2{\left(1+\beta\frac{GR_s}{r^3}\right)}^4\right],
\eea
where the trace of the logarithm of an area metric is defined as the trace of the matrix resulting from the flattened tensor as discussed extensively  in Ref. \cite{bianconi2024gravity}.
Here the Lagrangian $\mathcal{L}$ is  defined as long as ${\Delta}$ is positively defined, i.e., for 
\bea
r>r_0=\Big(2\beta{GR_s}\Big)^{1/3}={\Big(4\beta G^2M\Big)}^{1/3}.
\eea
We define the entropy of the Schwarzschild black hole as 
\bea
\mathcal{S}(R_s,\tau)&=&-\frac{1}{\ell_P^4}\int_0^{\tau} dt \int_{r_0}^{R_s} dr \int d\Omega \sqrt{-|g|} \mbox{Tr}\ln \tilde{\bf G}\tilde{g}^{-1}\nonumber \\
&=&-\frac{1}{\ell_P^4}\int_0^{\tau} dt \int_{r_0}^{R_s} dr \int d\Omega \sqrt{-|g|} \mbox{Tr}\ln {\Delta}.\nonumber
\eea 
Since the integral has the lower bound $r_0$, if follows that this entropy is defined only for $R_s>r_0$, which implies $R_s>R_0=\sqrt{2\beta G}$.
Moreover, we observe that when performing  the integral over time, in the expression for $\mathcal{S}$,  we consider the dimensional scale
\bea
\tau=\kappa^{-1}\tau^{\prime}=4GM\tau^{\prime},
\eea
where $\kappa^{-1}=4GM$ is the surface gravity. This way, we derive the explicit expression for the quantum relative entropy of the Schwarzschild black hole as a function of its radius $R_s$ and of $\tau^{\prime}$ given by 
\bea
\mathcal{S}(R_s,\tau^\prime)&=&\frac{32\pi M\tau^{\prime}}{3G}\left[3R_s^3\ln R_s^3+6\beta GR_s\ln\left(\frac{3}{2}\right)\right.\nonumber \\
&&-(R_s^3-2\beta GR_s)\ln(R_s^3-2\beta GR_s)\nonumber \\
&&\left.-2\left(R_s^3+\beta{GR_s}\right)\ln\left(R_s^3+\beta{GR_s}\right)\right].
\eea
In the limit $R_s\gg 1$, we find that  $\mathcal{S}(R_s,\tau^\prime)$ is linear in the Schwarzschild radius $R_s$, i.e.,
\bea
\mathcal{S}(R_s,\tau^\prime)={64\pi M\tau^{\prime}}\beta\ln\left(\frac{3}{2}\right) R_s,
\eea
and thus, for $\tau^{\prime}=\tau^{\prime}_1+\tau^{\prime}_2$, we find
\bea
S(R_s,\tau^{\prime})=S(R_s,\tau^{\prime}_1)+S(R_s,\tau^{\prime}_2),
\eea
and  for $R=R_{s,1}+R_{s,2}$ with $R_{s,i}\gg1 $, 
\bea
S(R_s,\tau^{\prime})=S(R_{s,1},\tau^{\prime})+S(R_{s,2},\tau^{\prime}).
\eea
In the above expression, we have considered $R_s$ as an independent variable from $M$. Let  us now impose that $R_s=2GM$  and consider the change of variables such that the entropy becomes a function of $M$ and $\tau^{\prime}$, i.e., $\mathcal{S}=\mathcal{S}(M,\tau^{\prime})$. For $R_s\gg1$, i.e., $M\gg1$, the black-hole entropy obeys the area law with 
\bea
\mathcal{S}(M,\tau^{\prime})\simeq \mathcal{S}_{A}=\mathcal{C}\frac{A}{4G},\label{SA}
\eea
where the area $A$ of the black hole is given by 
\bea
A=16\pi G^2 M^2,
\eea
and the multiplicative constant $\mathcal{C}$ is given by
\bea
\mathcal{C}=32 \ln (3/2)\beta\tau^\prime\simeq  \beta\tau^\prime\times 12.9749\ldots.
\eea
It follows that the quantum relative entropy $\mathcal{S}$ retains at the same time its information theory interpretation as a quantity that evaluates the local degree of freedom of the geometry, integrated over the volume of the black hole, while it can account for the emergence of the area law of the black-hole entropy.

\begin{figure}[!h!tb]
  \includegraphics[width=0.95\columnwidth]{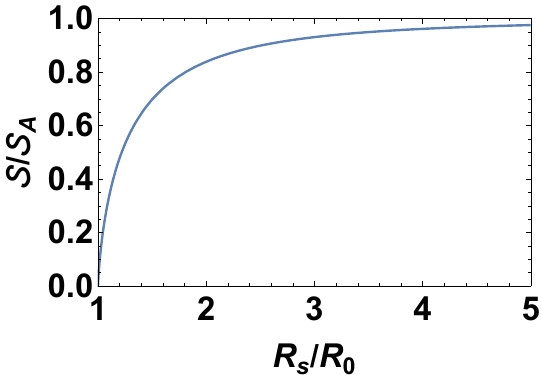}
  \caption{The quantum relative entropy of the Schwarzschild metric $\mathcal{S}$ divided by its asymptotic expression $\mathcal{S}_A=\mathcal{C}A/(4G)$, obeying the area law, is plotted as a function of the Schwarzschild radius $R_s$ for $G=1$.}
  \label{fig_entropy}
 \end{figure}

For  $0<R_s- R_0\ll 1$, we obtain
\bea
\mathcal{S}(M,\tau^{\prime})\simeq \frac{32\pi M \beta\tau^{\prime}}{3}(\mathcal{B}-4\ln({R_s}/{R_0}-1))(R_s-R_0),\nonumber 
\eea
with $\mathcal{B}=4-6\ln (3/2)$. Thus, $\mathcal{S}\to 0$ as $R\to R_0$.
We  define the temperature of the black hole as
\bea
\frac{1}{T}=\frac{\partial \mathcal{S}}{\partial M},
\eea
obtaining, in the limit $M\gg 1$,
\bea
T\to \frac{T_H}{\mathcal{C}},
\eea where $T_H$ is Hawking's temperature $T_H^{-1}=8\pi GM$. In the limit $R_s\to R_0=\sqrt{2\beta G}$ and $M\to \sqrt{\beta/(2G)})$, we obtain instead
\bea
T\simeq \left[-\frac{128\pi \beta R_0 \tau^{\prime}}{3}\ln \left(\frac{R_s}{R_0}-1\right)\right]^{-1}\to 0.
\eea
The quantum relative entropy of the Schwarzschild metric $\mathcal{S}$ divided by its asymptotic expression $\mathcal{S}_A$ given by Equation (\ref{SA}) is plotted in Figure \ref{fig_entropy} as a function of $R_s$ for $G=1$.

We note that if  entropic quantum gravity captures the true physics of gravitation, the quantum relative entropy of the Schwarzschild black hole only provides an approximation  for the entropy of physical black holes, valid in the limit for large Schwarzschild  radii, where the integral defining $\mathcal{S}$ is dominated by the terms of the small curvature.
Thus, the limit $R_s\simeq \sqrt{G}$ is the one in which the entropy of the Schwarzschild  metric  most deviates from the entropy of the black hole described by the entropic quantum gravity equation of~motion.

\section{Conclusions}
In conclusion, in this work, we considered the quantum relative entropy of a Schwarzschild  black hole. The quantum relative entropy is the central action in the entropic quantum gravity proposed in Ref. \cite{bianconi2024gravity}. It evaluates the quantum relative entropy between the metric associated with the considered manifold and the metric induced by the geometry and the matter field.
This metric depends on the curvature not only through the Ricci scalar and the Ricci tensor but also through the Riemann tensor.
In particular, it does not vanish for the Schwarzschild  black hole that has a non-vanishing Riemann tensor.
Here, we reinterpreted the Schwarzschild  black-hole metric in the light of the entropic quantum gravity approach proposed in Ref. \cite{bianconi2024gravity}. Although the Schwarzschild  metric is not the exact solution of the modified Einstein equations obtained from the entropic quantum gravity approach, rather only an approximate solution, here, we calculated its associated quantum relative entropy. We showed that despite the fact that the quantum relative entropy was  defined as the integral over the interior of the Schwarzschild  black hole, the quantum relative entropy obeyed the area law in the limit of a large Schwarzschild  radius.

This work can be expanded in several directions. On one hand, embracing the entropic quantum gravity approach will entail solving the modified Einstein equations for the corresponding black hole. On the other hand, it would be important to provide an interpretation of the quantum relative entropy in light of the second quantization of the theory. Both directions are likely to provide new quantum information insights into the entropic quantum gravity approach, which might hopefully  be testable experimentally.
\bibliographystyle{unsrt}
\bibliography{references}
\end{document}